\documentclass[a4paper, prb, twocolumn,floatfix, showpacs]{revtex4}

\usepackage{amsmath}
\usepackage{amsbsy}
\usepackage{amssymb}
\usepackage{graphicx}

\newcommand{\B}{\mathrm{S}}
\newcommand{\T}{\mathrm{T}}
\newcommand{\Gammaph}{N}
\renewcommand{\d}{\mathrm{d}}

\newcommand{\Linv}{L_0^{-1}}
\newcommand{\hc}{\text{h.c.}}
\newcommand{\Escale}{E_0}
\newcommand{\J}{\Lambda}
\newcommand{\Gammatransition}{\Gamma_\mathrm{on,heat}}

\begin{document}

\title{Cotunneling current through quantum dots with phonon-assisted spin-flip processes}
\author{J\"org Lehmann}
\email{joerg.lehmann@unibas.ch}
\author{Daniel Loss}
\affiliation{Department of Physics and Astronomy, University of Basel,
Klingelbergstrasse 82, CH-4056 Basel, Switzerland}
\date{\today}

\begin{abstract}
  We consider cotunneling through a quantum dot in the presence of
  spin-flip processes induced by the coupling to acoustic phonons of
  the surrounding. An expression for the phonon-assisted cotunneling
  current is derived by means of a generalized Schrieffer-Wolff
  transformation. The influence of the spin-phonon coupling on the
  heating of the dot is considered. The result is evaluated for the
  case of a parabolic semiconductor quantum dot with Rashba and
  Dresselhaus spin-orbit coupling and a method for the
  determination of the spin-phonon relaxation rate is proposed.
\end{abstract}

\pacs{
73.21.La, 
05.60.Gg, 
72.25.Rb, 
03.65.Yz  
}

\maketitle

\section{Introduction}

In the past years, prospective applications in spintronics and quantum
computation~\cite{Awschalom2002a,Zutic2004a,Nielsen2000a,Cerletti2005a}
have lead to a growing interest in the magnetic properties of
nanoscale systems. A variety of systems ranging from semiconductor
quantum dots over molecular
magnets~\cite{Sessoli1993a,Leuenberger2001a,Liang2002a,Meier2003a,Troiani2005a}
down to single atoms on surfaces have been studied in
detail.~\cite{Heinrich2004a,Zhao2005a} For the investigation of such systems,
transport measurements provide an excellent tool.  Recently, the
Zeeman splitting of two spin-levels has been measured via inelastic
tunneling spectroscopy, both in GaAs quantum dots~\cite{Kogan2004a}
and in single Mn atoms on a surface which have been addressed by an
STM tip.~\cite{Heinrich2004a} Using the same technique, the
singlet-triplet splitting in few-electron quantum dots has been
determined.~\cite{Zumbuehl2004a}

For the applications mentioned at the beginning it is crucial that
effects like relaxation and the loss of quantum coherence resulting
from the coupling of the nanosystem to its surrounding environment are
sufficiently weak. In particular, a detailed knowledge about these
effects is a prerequisite for the assessment of the suitability of a
specific system for application purposes. In the case of the electron
spin in semiconductor quantum dots, relaxation at low magnetic fields
is dominated by the hyperfine coupling to the
nuclei,~\cite{Erlingsson2002a,Khaetskii2002a,Coish2004a} while at
higher fields the spin-orbit interaction induced coupling to
acoustical phonons is most relevant.~\cite{Golovach2004a,Bulaev2005a}
It turns out that both effects are sufficiently weak, and
correspondingly long spin-relaxation times have been
reported.~\cite{Fujisawa2002a,Elzerman2004a, Kroutvar2004a} For the
single-atom experiments reported in Ref.~\onlinecite{Heinrich2004a},
the strength of different coupling mechanisms is not yet clear,
however, and the measurement schemes which were used in the
semiconductor case are not directly applicable.

In the present work, we study to what extent inelastic cotunneling
spectroscopy can yield information about intrinsic spin-flip processes
due to a coupling to bosonic degrees of the freedom in the surrounding.
We put forward a general approach for the calculation of the cotunneling
current~\cite{Averin1990a, Golovach2004b} in the presence of such
spin-flip processes. Subsequently, we apply the general formalism to the
case of cotunneling across GaAs/AlGaAs semiconductor quantum dots and
propose a method for the determination of the spin-relaxation rate.

The outline of the present work is as follows. In
Sec.~\ref{sec:model}, we start by introducing the model for the
quantum dot coupled via a tunneling Hamiltonian to two fermionic leads
and via a spin-phonon coupling to a bosonic environment.  In the next
section, we put forward a generalized Schrieffer-Wolff transformation,
which, in the limit of a weak spin-phonon coupling, allows one to
eliminate to lowest order the dot-lead coupling. We then derive, in
Sec.~\ref{sec:current}, expressions for the elastic and inelastic
cotunneling current in the presence of the spin-phonon coupling. For
the evaluation of the current, the knowledge of the occupation
probabilities of the spin-states on the dot is required.  Their
dynamics is governed by a master equation, which is discussed in
Sec.~\ref{sec:master-equation}. The experimentally most relevant
quantity, the differential conductance as a function of the bias
voltage, is evaluated in Sec.~\ref{sec:conductance}. In
Sec.~\ref{sec:kogan}, we apply the general result to a realistic
model describing a recent experiment by Kogan et al.
(Ref.~\onlinecite{Kogan2004a}). Sec.~VIII briefly discusses the
low-temperature behavior. The final conclusions are presented in
Sec.~\ref{sec:conclusions}.

\section{Model}
\label{sec:model}

\begin{figure}[ht]
  \centerline{\includegraphics[width=0.65\linewidth]{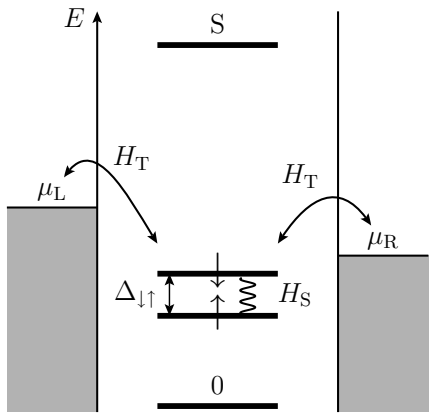}}
  \caption{Sketch of the quantum dot coupled via the tunneling
    Hamiltonian $H_\T$ to two leads at electro-chemical potentials
    $\mu_\mathrm{L}$ and $\mu_\mathrm{R}$ and to a phonon bath via the
    spin-phonon coupling Hamiltonian $H_\B$. The dot can be either
    empty ($0$), in one of the spin-directions $\uparrow$ and
    $\downarrow$, which are split by the Zeeman energy
    $\Delta_{\downarrow\uparrow}$, or in the singlet state S.}
  \label{fig:model}
\end{figure}

We consider transport across a quantum dot in a spin-$1/2$
ground-state contacted to two leads in the presence of an intrinsic
spin-flip mechanism due to the coupling of the dot to a phonon bath
(see Fig.~\ref{fig:model}).
The total system consisting of dot, leads and phonons is described by
the Hamiltonian
\begin{equation}
  H = H_0 + H_\T + H_\B \,.
\end{equation}
Here, $H_0$ is the Hamiltonian of the isolated dot, the ideal leads
$\ell=\mathrm{L}$, $\mathrm{R}$ and the phonons, i.e.,
\begin{equation}
  \label{eq:H0}
  H_0 = \sum_{\sigma=\uparrow,\downarrow} E_\sigma n_\sigma +
  U n_\uparrow n_\downarrow
  +
  \sum_{\ell \mathbf{k} \sigma} \epsilon_{\ell \mathbf{k}} n_{\ell \mathbf{k}\sigma}
  +
  \sum_\mathbf{q} \hbar\omega_\mathbf{q} n_\mathbf{q}
  \,.
\end{equation}
The number operator of the dot spin is given by $n_\sigma =
d^\dagger_\sigma d_\sigma$, where $d_\sigma$ ($d^\dagger_\sigma$)
destroys (creates) an electron with spin $\sigma=\uparrow,
\downarrow$ on the dot.  An externally applied magnetic field $B_z$ in
z-direction leads to a Zeeman splitting $\Delta_{\downarrow\uparrow} =
E_\downarrow - E_\uparrow = - g \mu_\mathrm{B} B_z$ of the
single-particle dot levels.  Furthermore, due to electron-electron
interaction, the simultaneous occupation of the dot by two electrons
costs the additional energy~$U$.
The leads $\ell=\mathrm{L,R}$ are modeled as Fermi liquids, i.e., by non-interacting
quasiparticles with energy $\epsilon_{\ell \mathbf{k}}$, where $\mathbf{k}$ is
the wave-vector and $\sigma$ the spin. The number operator $n_{\ell \mathbf{k}
  \sigma}=c^\dagger_{\ell \mathbf{k}\sigma}\,c_{\ell \mathbf{k}\sigma}$ for the
corresponding state is defined in terms of the fermionic creation and
destruction operators $c^\dagger_{\ell \mathbf{k}\sigma}$ and $c_{\ell
  \mathbf{k}\sigma}$, respectively. Each lead is filled up to an
electro-chemical potential~$\mu_\ell$, and thus the mean occupation numbers
obey the Fermi distributions
$f_\ell(\epsilon) = \left\{\exp\left[(\epsilon-\mu_\ell)/k_\mathrm{B}T\right]+1\right\}^{-1}$.
Here, an externally applied
bias voltage~$V$ maps to a difference
$\mu_\mathrm{L}-\mu_\mathrm{R}=eV$. In the present work, we consider
transport across a singly-occupied dot in the cotunneling regime.
This imposes the condition $E_\sigma\ll \mu_\ell \ll E_\sigma + U$ for
$\sigma = \uparrow, \downarrow$ and $\ell=\mathrm{L},\mathrm{R}$.  The
last term in Eq.~\eqref{eq:H0} describes the phonons of the
environment, which consist of a collection of modes characterized by a
wave-vector $\mathbf{q}$ with occupation numbers $n_\mathbf{q} =
a^\dagger_\mathbf{q} a_\mathbf{q}$. Here, the bosonic operators
$a^\dagger_\mathbf{q}$ ($a_\mathbf{q}$) create (destroy) a phonon in
the mode~$\mathbf{q}$. Note that for reasons of notational simplicity,
a possible branch index has been suppressed here, but can be added in
the end results. At equilibrium, the phonons are described by the
Bose-distribution $n_\mathrm{B}(\epsilon)=[\exp(\epsilon/k_\mathrm{B}T)-1]^{-1}$.

The coupling of the dot to the leads is modeled by the tunneling Hamiltonian
\begin{equation}
  H_\T = \sum_{\ell \mathbf{k} \sigma} T_{\ell \mathbf{k}}\, c^\dagger_{\ell \mathbf{k} \sigma}
  d_{\sigma} + \hc\,,
\end{equation}
where, for reasons of simplicity, we have assumed spin-independent
tunnel matrix elements $T_{\ell \mathbf{k}}$.
For later use, we introduce the corresponding tunneling rates (per spin-direction)
\begin{equation}
  \label{eq:gammal}
  \Gamma_\ell(\epsilon) =
  \frac{2\pi}{\hbar} \sum_\mathrm{k} |T_{\ell \mathbf{k}}|^2\, \delta(\epsilon-\epsilon_{\ell \mathbf{k}})\,.
\end{equation}
So far, except for the presence of two instead of one lead, the
Hamiltonian corresponds to the one of the single-site Anderson
model,~\cite{Anderson1961a} which has been extensively studied in the
literature.~\cite{Hewson1993a}

Finally, we assume that the dot spin interacts with the phonon system
via a spin-phonon coupling of 
the form
\begin{equation}
  \label{eq:HB}
  H_\B = \sum_{\mathbf{q}}
  (M_{\mathbf{q},x} \, \sigma_x + M_{\mathbf{q},y} \, \sigma_y) (a_\mathbf{q} + a_{-\mathbf{q}}^\dagger)\,.
\end{equation}
In order to guarantee the hermiticity of the coupling Hamiltonian, the
coefficients have to fulfil $M_{\mathbf{q},i} =
M_{-\mathbf{q},i}^\ast$ for all $\mathbf{q}$ and $i=x,y$. Furthermore,
we have introduced the spin operators $\sigma_x = d^\dagger_\downarrow
d_\uparrow + d^\dagger_\uparrow d_\downarrow $ and $\sigma_y = i
(d^\dagger_\downarrow d_\uparrow - d^\dagger_\uparrow d_\downarrow )$.
Again, we can characterize the coupling by a spectral density, which
is defined as
\begin{equation}
  \label{eq:d}
  D(\omega) = \frac{\pi}{\hbar} \sum_{\mathbf{q}} |M_{\mathbf{q},x}
  + i M_{\mathbf{q},y}|^2 \, \delta(\omega-\omega_\mathbf{q})\,.
\end{equation}
Note that the coupling does not contain dephasing terms proportional
to $\sigma_z= d^\dagger_\uparrow d_\uparrow - d^\dagger_\downarrow
d_\downarrow$ or to the total number of electrons
$n=n_\uparrow+n_\downarrow$ on the dot.  For the electron spin in a
quantum dot this is justified to a very good approximation, as was
shown in Ref.~\onlinecite{Golovach2004a}. In this regard, the present
model differs, e.g., from the so-called Anderson-Holstein
model,~\cite{Holstein1959a,Anderson1975a} which only contains
coupling terms of this kind.

\section{Schrieffer-Wolff transformation}
\label{sec:schriefferwolff}

Since we are interested in transport in the cotunneling regime where
the dot occupancy is only changed virtually, it is advantageous to
eliminate to lowest order the tunneling Hamiltonian $H_\T$, which
changes the number of electrons on the dot by one. A standard method
to achieve this goal is the so-called Schrieffer-Wolff
transformation,~\cite{Schrieffer1966a} which we generalize to the case
where a spin-flip coupling $H_\B$ is present. Accordingly, we perform
the canonical transformation to the Hamiltonian
\begin{equation}
  \bar{H} = e^S H e^{-S} = H_0 + H_\B +
  \frac{1}{2} [S, H_\T] + \mathcal{O}(H_\T^3)\,,
\end{equation}
where the generator~$S$ of the transformation has to fulfil the
condition $[S, H_0+H_\B] + H_\T = 0$. In order to obtain an explicit
expression for the generator~$S$, we now use that, as has been
discussed in the Introduction, in all relevant cases, the spin-flip
coupling is weak and can be treated perturbatively.  Hence, we can
expand
\begin{equation}
S = \Linv H_\T - \Linv [\Linv H_\T,
H_\B] + \mathcal{O}(H_\B^2)\,.
\end{equation}
Here, $L_0$ is the Liouvillian of the decoupled dot, leads and phonon
system.~\footnote{The Liouvillian superoperator $L_0$ has vanishing
  diagonal matrix elements in the basis $|n\rangle\langle n'|$, where
  $|n\rangle$ are the eigenstates of the Hamiltonian $H_0$. In order
  to obtain an invertible operator, one has to add an
  infinitesimal imaginary part. $L_0^{-1}$ has to be understood in
  this sense. The resulting infinitesimal imaginary parts in the
  denominators of Eq.~\eqref{eq:H0dir-exch} and the like, are,
  however, not relevant for the following discussion.} We, thus,
arrive at the transformed Hamiltonian
\begin{equation}
  \label{eq:SWT}
  \bar H \approx H_0 + H_\B + \frac{1}{2} [\Linv H_\T, H_\T]
  + \frac{1}{2} [\Linv [\Linv H_\T, H_\B], H_\T]\,.
\end{equation}
The first commutator corresponds to the standard Schrieffer-Wolff
transformation~\cite{Schrieffer1966a} and yields a contribution
\begin{equation}
  \label{eq:4}
 \frac{1}{2} [\Linv H_\T, H_\T]
 =
 H_\mathrm{dir,ex}
 + 
 H_\mathrm{ren} 
 +
 H_\mathrm{2e}\,.
\end{equation}
Here, the first term contains the direct and exchange interaction
between the dot and the leads
\begin{multline}
  \label{eq:H0dir-exch}
  H_\mathrm{dir,ex} = 
  \frac{1}{2}
   \sum_{\ell \ell' \mathbf{k} \mathbf{k}' \sigma \alpha}
   \frac{T_{\ell \mathbf{k}}
     T_{\ell' \mathbf{k}'}^\ast}{\epsilon_{\ell \mathbf{k}}-E_\sigma^\alpha}
   \big(
   n_{\bar\sigma}^\alpha \, c_{\ell \mathbf{k} \sigma}^\dagger c_{\ell' \mathbf{k}'\sigma}\\
   -
   \alpha\,
   d_{\bar\sigma}^\dagger \, d_\sigma \, c_{\ell \mathbf{k} \sigma}^\dagger\,  c_{\ell' \mathbf{k}'\bar\sigma}
   \big)
   + \hc
\end{multline}
We denote flipped spins by $\bar\uparrow = \downarrow$ and vice versa.
Here and in the following, the sums over $\alpha$ run over $\alpha=\pm
1$. Furthermore, we have introduced the abbreviations $n^\alpha_\sigma
= (1-\alpha)/2 + \alpha \,n_\sigma$ and $E_\sigma^\alpha = E_\sigma +
[(\alpha+1)/2] U$. The next contribution to Eq.~\eqref{eq:4} is the
renormalization of the dot Hamiltonian $H_0$,
\begin{equation}
  \label{eq:H0ren}
   H_\mathrm{ren} = 
   \sum_{\ell \mathbf{k} \sigma \alpha} \frac{|T_{\ell \mathbf{k}}|^2}{\epsilon_{\ell
       \mathbf{k}}- E_\sigma^\alpha} \, n_{\bar\sigma}^\alpha \,n_\sigma\,.
\end{equation}
As this contribution can be absorbed in the definition of $H_0$, it
will be ignored in the following.
The last term, which describes processes that change the electron number on
the dot by two, reads
\begin{equation}
  \label{eq:H02e}
   H_\mathrm{2e} = 
   \frac{1}{2}
   \sum_{\ell \ell' \mathbf{k} \mathbf{k}' \sigma \alpha}
   \alpha\,
   \frac{T_{\ell \mathbf{k}} T_{\ell' \mathbf{k}'}^\ast}{\epsilon_{\ell
       \mathbf{k}}-E_\sigma^\alpha}\,
   d_{\bar\sigma} d_\sigma \, c^\dagger_{\ell' \mathbf{k}' \bar\sigma} c^\dagger_{\ell
   \mathbf{k} \sigma}
 + \hc
\end{equation}
Later, we shall focus on the sector of the Hilbert space where the dot
is singly occupied, and hence $H_\mathrm{2e}$ can be neglected, as well.

Similarly, we can decompose the phonon-mediated contributions to the
transformed Hamiltonian~\eqref{eq:SWT} into
\begin{equation}
  \label{eq:6}
  \frac{1}{2} [\Linv [\Linv H_\T, H_\B], H_\T] =
  H^\mathrm{ph}_\mathrm{dir,ex}
  + 
  H^\mathrm{ph}_\mathrm{ren} 
  +
  H^\mathrm{ph}_\mathrm{2e}\,.
\end{equation}
Here, the relevant contribution containing direct and exchange terms reads
\begin{equation}
  \label{eq:Hphdir-exch}
  \begin{split}
    H^\mathrm{ph}_\mathrm{dir,ex} =
    \frac{1}{2} &
      \sum_{\ell \ell' \mathbf{k} \mathbf{k}' \sigma \alpha}
      T_{\ell \mathbf{k}} \, T_{\ell' \mathbf{k}'}^\ast\,
      X_{\ell \mathbf{k} \sigma \alpha}
      \\
      & \times 
      \left( 
        n^\alpha_{\sigma}\,
      c^\dagger_{\ell \mathbf{k} \sigma} c_{\ell' \mathbf{k}' \bar \sigma}
      -
      \alpha
      d^\dagger_{\sigma} d_{\bar \sigma}\,
      c^\dagger_{\ell \mathbf{k} \sigma} c_{\ell' \mathbf{k}' \sigma}
      \right) + \mathrm{h.c.}
  \end{split}
\end{equation}
where we have introduced the phonon operator
\begin{multline}
  X_{\ell \mathbf{k} \sigma \alpha} =
  \sum_{\mathbf{q}, i=x,y}
  \frac{M_{\mathbf{q},i}\,\sigma_i^{\sigma\bar\sigma}}
  {\epsilon_{\ell \mathbf{k}} -   E_\sigma^\alpha}
  \bigg(
    \frac{a_\mathbf{q}}{\epsilon_{\ell \mathbf{k}} - E_{\bar\sigma}^\alpha -
      \hbar\omega_\mathbf{q}}\\
    +
    \frac{a^\dagger_{-\mathbf{q}}}{\epsilon_{\ell \mathbf{k}} - E_{\bar\sigma}^\alpha +
      \hbar\omega_\mathbf{q}}
  \bigg)  \,.
\end{multline}
Furthermore, the transformation leads to a renormalization of the
spin-phonon coupling
\begin{equation}
  \label{eq:HphBren}
  H^\mathrm{ph}_\mathrm{ren} 
  =
  -
  \frac{1}{2}
  \sum_{\ell  \mathbf{k}  \sigma}
  |T_{\ell \mathbf{k}}|^2
  X_{\ell \mathbf{k} \sigma -}
  d^\dagger_\sigma d_{\bar\sigma} + \hc 
\end{equation}
Finally, a two-electron tunneling term 
\begin{equation}
  \label{eq:Hph2e}
  H^\mathrm{ph}_\mathrm{2e} =
  \frac{1}{2}
  \sum_{\ell \ell' \mathbf{k} \mathbf{k}' \sigma \alpha}\!\!\!\!\!\!
  \alpha\,
  T_{\ell \mathbf{k}} T_{\ell' \mathbf{k}'}\, 
  X_{\ell \mathbf{k} \sigma \alpha}\,
  d_{\sigma} \, d_{\bar\sigma} \,
  c_{\ell' \mathbf{k}'\sigma}^\dagger c_{\ell \mathbf{k} \sigma}^\dagger
  + 
  \hc
\end{equation}
is generated, which, for the same reasons as above, will be
disregarded together with the renormalization term~\eqref{eq:HphBren}.

\section{Current}
\label{sec:current}

The mean current $I_\ell(t)$ across contact
$\ell=\mathrm{L},\mathrm{R}$ is given by the expectation value of the
time-derivative of the total number of electrons $N_\ell =
\sum_{\mathbf{k}\sigma} n_{\ell \mathbf{k} \sigma}$ in lead $\ell$ multiplied
by the electron charge $-e$, i.e.,
\begin{equation}
I_\ell(t) = -e \langle \dot{N}_\ell \rangle = -\frac{i e}
{\hbar}\langle [\bar H, N_\ell] \rangle\,.
\end{equation}
For the evaluation of this expectation value, we switch to the
interaction picture to obtain \cite{Mahan2000a,Kohler2005a}
\begin{equation} 
  \label{eq:meancurrent}
  I_\ell (t)=
  \frac{i e}{\hbar}
  \int\limits_0^{t-t_0}
  \!\d\tau\,
  \langle
  [\dot{\tilde{N}}_\ell(t-t_0),
  \tilde{H}_1(t-\tau-t_0)]\,
  \rangle
  \,.
\end{equation}
Here, the tilde denotes interaction picture operators with respect to
$H_0$ and $H_1:=\bar H-H_0$ is the coupling part of the effective
Hamiltonian and the expectation value has to be taken at time
$t-\tau-t_0$.

In the long-time limit $t\gg t_0$, the mean value of the current becomes
time-independent and we find that due to charge conservation
$I_\mathrm{L}=-I_\mathrm{R}$. Hence, we can write the total current as
$I = I_\mathrm{RL} - I_\mathrm{LR}$, where $I_{\ell'\ell}$ is the
contribution of the electrons flowing from lead $\ell$ into
lead~$\ell'$. 
In the
cotunneling regime, where $E_\sigma\ll \mu_\ell \ll E_\sigma + U$ for
$\sigma = \uparrow, \downarrow$ and $\ell=\mathrm{L},\mathrm{R}$, the
dot is, up to exponentially small corrections, always occupied by a
single electron and hence is characterized by the occupation
probabilities $p_\sigma = \langle n_\sigma \rangle$ with $p_\uparrow +
p_\downarrow = 1$. The currents $I_{\ell'\ell}$ can then be written as~\cite{Sukhorukov2001,Golovach2004b}
\begin{equation}
\label{eq:current2}
I_{\ell'\ell} = e \sum_{\sigma\sigma'} \, W_{\ell'\sigma'\ell\sigma}\,
p_\sigma \,,
\end{equation}
where $W_{\ell'\sigma'\ell\sigma}$ is the rate for an
electron tunneling from lead~$\ell$ into lead~$\ell'$ when the dot was
initially in the state~$\sigma$ and is in the state~$\sigma'$ afterwards. By convention, a process with
$\sigma'=\sigma$ is called \textit{elastic} cotunneling, while the
case with $\sigma'=\bar\sigma$ is referred to as \textit{inelastic}
cotunneling.~\cite{Averin1990a}

The contribution due to the cotunneling without participation of
phonons to the rates~$W_{\ell'\sigma'\ell\sigma}$ will be designated
by $W^{(0)}_{\ell'\sigma'\ell\sigma}$. In addition, the phonon-assisted
cotunneling process contributes with rates
$W^{(1)}_{\ell'\sigma'\ell\sigma}$ to the total rates
$W_{\ell'\sigma'\ell\sigma} =
W^{(0)}_{\ell'\sigma'\ell\sigma}+W^{(1)}_{\ell'\sigma'\ell\sigma}$. In
the next two subsection, we discuss these two contributions in detail.
Note that we thereby restrict ourselves to temperatures above the
Kondo temperature, where correlations between the dot spin and the
collective spin of the electrons in the lead do not play a role. We
shall come back to a brief discussion of the low-temperature behavior
in Sec.~\ref{sec:pms}.

\subsection{Elastic and inelastic cotunneling}
\label{sec:elast-inel-cotunn}

Evaluating the current expression~\eqref{eq:meancurrent} for the direct and
exchange terms~\eqref{eq:H0dir-exch}, we arrive at the elastic
cotunneling rate
\begin{equation}
  \label{eq:Wcotel}
  \begin{split}
    W^{(0)}_{\ell'\sigma\ell\sigma} =
    \frac{\hbar}{2\pi}
    \int\d\epsilon &\,
    \Gamma_{\ell'}(\epsilon)
    \,
    \Gamma_\ell(\epsilon)\,
    \J^{(0),\mathrm{el}}_{\sigma}(\epsilon)
    f_\ell(\epsilon) \left[1 - f_{\ell'}(\epsilon) \right]
  \end{split}
\end{equation}
with $\J^{(0),\mathrm{el}}_\sigma(\epsilon) =
\left(E_{\bar\sigma}+U-\epsilon\right)^{-2} +
\left(E_{\sigma}-\epsilon\right)^{-2}$. Similarly, the inelastic
cotunneling rate is given by
\begin{equation}
  \label{eq:Wcotinel}
  \begin{split}
    W^{(0)}_{\ell'\bar\sigma\ell\sigma}  =
    \frac{\hbar}{2\pi}
    \int\d\epsilon &\,
    \Gamma_{\ell'}(\epsilon-\Delta_{\bar\sigma \sigma})
    \,
    \Gamma_{\ell}(\epsilon)\,
    \J^{(0),\mathrm{inel}}_{\sigma}(\epsilon)
    \\&\times
    f_{\ell}(\epsilon)
    \left[
    1 - f_{{\ell'}}(\epsilon- \Delta_{\bar\sigma \sigma})
    \right]\,,
  \end{split}
\end{equation}
where $\J^{(0),\mathrm{inel}}_\sigma(\epsilon) =
\big[\left(E_{\bar\sigma}+U-\epsilon\right)^{-1} -
\left(E_{\bar\sigma}-\epsilon\right)^{-1}\big]^2$. In the inelastic
case, the energy~$\Delta_{\bar\sigma \sigma}$ is deposited in or
extracted from the dot. Correspondingly, the initial energy~$\epsilon$
of the tunneling electron coming from lead~$\ell$ differs by the
same amount from the final energy of the electron in lead~$\ell'$.
Inelastic cotunneling is, thus, strongly suppressed for temperatures
and external voltages much smaller than~$\Delta_{\bar\sigma \sigma}$.

To a very good approximation, the energy dependence of the coupling
rates $\Gamma_\ell(\epsilon)$ can be ignored in the energy range
contributing to the integrals in the rate
expressions~\eqref{eq:Wcotel} and \eqref{eq:Wcotinel}.  Furthermore,
near one of the resonances, e.g., for $E_\sigma \lesssim \mu_\ell \ll
E_{\bar\sigma}+U$, we can expand the dominant energy denominator of
$\J^{(0),\mathrm{el}}_\sigma(\epsilon)$ and
$\J^{(0),\mathrm{inel}}_\sigma(\epsilon)$, respectively, and carry out
the energy integration.~\footnote{By expanding the energy
  denominators, we remove divergences at $\epsilon=E_\sigma$ and
  $\epsilon=E_\sigma+U$, which are, however, exponentially suppressed
  in the cotunneling regime.} This yields the approximate rates
\begin{multline}
  \label{eq:Icot2}
    W^{(0)}_{\ell'\sigma'\ell\sigma} \approx
    \frac{\hbar
      \Gamma_\ell\Gamma_{\ell'}
      }{2\pi}
    \frac{1}{\Delta_{\ell'\sigma'\ell\sigma}^2}
    \bigg\{1 +         \frac{1}{4\Delta_{\ell'\sigma'\ell\sigma}^2} 
    \\\times
    \left[ 4\pi^2
      (k_\mathrm{B} T)^2 + \left(\mu_{\ell'\ell}+\Delta_{\sigma'\sigma}\right)^2\right]\bigg\}
    \Theta(-\mu_{\ell'\ell}-\Delta_{\sigma'\sigma})
\end{multline}
where we have introduced the function
\begin{equation}
  \label{eq:11}
  \Theta(\epsilon)=\frac{\epsilon}{1-\exp(-\epsilon/k_\mathrm{B}T)}
\end{equation}
as well as the energies
\begin{equation}
  \label{eq:Delta}
  \Delta_{\ell'\sigma'\ell\sigma} =
  \frac{E_{\sigma'}-\mu_{\ell'} + E_\sigma-\mu_\ell }{2}
\end{equation}
and
$\mu_{\ell'\ell}=\mu_{\ell'}-\mu_\ell$. Here, terms of the order
$\mathcal{O}\big(\left[\max(k_\mathrm{B}T,
|\mu_{\ell'\ell}+\Delta_{\sigma'\sigma}|)/\Delta_{\ell'\sigma'\ell\sigma}\right]^4\big)$
have been neglected in the curly brackets.

\subsection{Phonon-assisted elastic and inelastic cotunneling}

We now come to the current contributions due to the presence of the
spin-phonon coupling. Using $H^\mathrm{ph}_\mathrm{dir,ex}$ from
Eq.~\eqref{eq:Hphdir-exch}, we obtain after a straightforward but
lengthy calculation the rates for the phonon-assisted elastic
cotunneling
\begin{equation}
  \label{eq:Icotphel}
  \begin{split}
    W^{(1)}_{\ell'\sigma\ell\sigma} =
    \frac{\hbar}{8\pi^2}
    \int\!\d\epsilon' &\,\Gamma_{\ell'}(\epsilon')
    \int\!\d\epsilon \,\Gamma_{\ell}(\epsilon)\,
    \J^{(1)}_\mathrm{el}(\epsilon,\epsilon')\\
    & \times
    \Gammaph(\epsilon-\epsilon')\,
    f_{\ell}(\epsilon) \left[1-f_{\ell'}(\epsilon')\right]\,.
  \end{split}
\end{equation}
The corresponding energy denominator 
\begin{multline}
  \label{eq:Jcotphel}
\J^{(1)}_\mathrm{el}(\epsilon, \epsilon') =
\sum_\sigma
\bigg[
\frac{1}{(E_\sigma + U - \epsilon)\,(E_{\bar\sigma} + U - \epsilon')}\\
+
\frac{1}{(E_\sigma  - \epsilon)\,(E_{\bar\sigma} - \epsilon')}
\bigg]^2
\end{multline}
now depends on both the energy before and after the phonon
emission and/or absorption. Note that as explained above, we use the term
``elastic'' only to indicate that the dot state before and after the
cotunneling process is identical. From the point of view of the
electron system, the phonon-mediated cotunneling process is, of
course, no longer elastic.  Rather, the electrons either emit the
energy $\epsilon-\epsilon'$ due to the stimulated and spontaneous
emission of phonons or absorb the energy
$\epsilon'-\epsilon$ from the phonon system. Both processes are
captured by the function

\begin{equation}
\Gammaph(\Delta\epsilon) =
D(\Delta\epsilon/\hbar) \, \left[n_\mathrm{B}(\Delta\epsilon)+1\right]
+ D(-\Delta\epsilon/\hbar) \, n_\mathrm{B}(-\Delta\epsilon)\,.  
\end{equation}

Similarly, we obtain the ``inelastic'' phonon-mediated cotunneling
rates
\begin{equation}
  \label{eq:Wcotphinel}
  \begin{split}
    W^{(1)}_{\ell'\bar\sigma\ell\sigma} =
    \frac{\hbar}{8\pi^2}
    &
    \int\!\d\epsilon' \,\Gamma_{\ell'}(\epsilon')
    \int\!\d\epsilon \,\Gamma_{\ell}(\epsilon)\,
    \J^{(1)}_\mathrm{inel}(\epsilon,\epsilon')\\
    &
    \times
    \Gammaph(\epsilon-\epsilon'-\Delta_{\bar \sigma \sigma})\,
    f_{\ell}(\epsilon) \left[1-f_{\ell'}(\epsilon')\right]
  \end{split}
\end{equation}
with 
\begin{multline}
  \label{eq:Jcotphinel}
\J^{(1)}_\mathrm{inel}(\epsilon, \epsilon') = \sum_\sigma\bigg[
\frac{1}{(E_\sigma + U - \epsilon) (E_\sigma + U - \epsilon')}\\
-
\frac{1}{(E_\sigma - \epsilon) (E_\sigma - \epsilon')}\bigg]^2\,.
\end{multline}

For the further evaluation of the rate formulas we again assume
energy-independent dot-lead couplings $\Gamma_\ell$.  Since the energy denominators in
Eqs.~\eqref{eq:Jcotphel} and \eqref{eq:Jcotphinel} are already of
fourth order in $\Delta_{\sigma\mu}$, we can neglect their energy
dependence around $\epsilon=\epsilon'=\mu$ to get a result valid to the same
order as in Eq.~\eqref{eq:Icot2}. Carrying
out one of the energy integrals, we obtain the approximate result
for both rates
\begin{multline}
  \label{eq:Wcotphinel2}
    W^{(1)}_{\ell'\sigma'\ell\sigma} \approx
    \frac{\hbar\Gamma_{\ell}\Gamma_{\ell'}}{8\pi^2}
    \left(
    \frac{1}{\Delta_{\bar\sigma'\ell}^2\, \Delta_{\sigma\ell'}^2}
    +
    \frac{1}{\Delta_{\sigma'\ell}^2 \,\Delta_{\bar\sigma\ell'}^2}
  \right)
    \\\times
    \int\!\d\Delta\epsilon \,
    \Gammaph(\Delta\epsilon)\,
    \Theta(-\mu_{\ell'\ell}-\Delta_{\sigma'\sigma}- \Delta\epsilon)
\end{multline}
with the energy differences $\Delta_{\sigma\ell} = E_\sigma -
\mu_\ell$.
We now consider a spectral density of the spin-phonon coupling of the
form
\begin{equation}
  \label{eq:spectraldens}
  D(\omega\ge 0) = \gamma \hbar \, \omega_\mathrm{c}^{1-s} \, \omega^s
\end{equation}
with the dimensionless spin-phonon coupling parameter~$\gamma$, the
cut-off frequency~$\omega_\mathrm{c}$ and an exponent~$s$, which
depends on the nature of the coupling.  In order for our
master-equation approach to be valid, we have to restrict ourselves to
exponents $s\ge1$. The exponent $s=1$ corresponds to the so-called
Ohmic case, and below, we will find $s=3$ or $s=5$ for a spin-orbit
mediated coupling to phonons.  Inserting the spectral
density~\eqref{eq:spectraldens} into Eq.~\eqref{eq:Wcotphinel2}, we
can write the rates as
\begin{multline}
  \label{eq:Wcotphinel3}
    W^{(1)}_{\ell'\sigma'\ell\sigma} \approx
    \gamma(\hbar\omega_\mathrm{c})^{1-s}
    \frac{\hbar\Gamma_{\ell}\Gamma_{\ell'}}{8\pi^2}
    \left(
    \frac{1}{\Delta_{\bar\sigma'\ell}^2\, \Delta_{\sigma\ell'}^2}
    +
    \frac{1}{\Delta_{\sigma'\ell}^2\,
      \Delta_{\bar\sigma\ell'}^2}
  \right)
  \\\times
    (k_\mathrm{B}T)^{1+s}
  J_s\left(-\left(\mu_{\ell'\ell}+\Delta_{\sigma'\sigma}\right)/k_\mathrm{B}T\right)
  \Theta\left(-\mu_{\ell'\ell}-\Delta_{\sigma'\sigma}\right)
\end{multline}
with 
\begin{equation}
  \label{eq:7}
  J_s(a) = 
  \frac{1-e^{-a}}{a}
  \int_0^\infty \!\!\d x \,
  \mathop\mathrm{sgn}\left( \ln x\right)
  \frac{|\ln x|^s}{x-1}
  \frac{\ln x + a}{x-e^{-a}}\,.
\end{equation}
In the interesting cases of $s=1$, $3$ and $5$, we find for the integral
\begin{align}
  \label{eq:5}
  J_1(a) & = \frac{1}{6}\left(4\pi^2+a^2\right)\\
  J_3(a)  & =
  \frac{1}{60}\left(4\pi^2+a^2\right)\left(8\pi^2+3a^2\right)\,,
  \\
  J_5(a)  & =
  \frac{1}{42}
  \left(2\pi^2+a^2\right)  
  \left(4\pi^2+a^2\right)
  \left(8\pi^2+a^2\right)\,.
\end{align}
In the general case, one can readily show that $J_s(a)$ is an even
function, i.e., $J_s(a)=J_s(-a)$ for all $s$ and $a$.

\section{Master equation}
\label{sec:master-equation}

For the evaluation of the current expressions given in the previous
section, we still need the occupation probabilities $p_\sigma$. Their
dynamics is governed by the master equation~\cite{Blum1996a} 
\begin{equation}
  \label{eq:mastereq}
  \dot{p}_\sigma = -W_{\bar\sigma\sigma} \, p_\sigma +
  W_{\sigma\bar\sigma} \, p_{\bar\sigma}
\end{equation}
with the total rates $W_{\bar\sigma\sigma} =
W^\mathrm{cot}_{\bar\sigma\sigma} +
W^\mathrm{flip}_{\bar\sigma\sigma}$.  Here,
$W^\mathrm{cot}_{\bar\sigma\sigma} = \sum_{\ell\ell'}
W_{\ell'\bar\sigma\ell\sigma}$ is the spin-flip rate due to inelastic
cotunneling, which, in contrast to the current
formula~\eqref{eq:current2}, also contains processes involving a
single lead only.  Furthermore, spin-flip processes due to the
spin-phonon coupling~(\ref{eq:HB}) have to be taken into account in
the master equation~\eqref{eq:mastereq}.  These processes lead to the
additional rate
\begin{equation}
  \label{eq:flip}
 W^\mathrm{flip}_{\bar\sigma\sigma}
 = \frac{2}{\hbar}\,
 \Gammaph(\Delta_{\sigma\bar\sigma})\,.
\end{equation}
In the stationary limit, the solution of the master equation is given
by $p_\sigma =
\left[1+W_{\bar\sigma\sigma}/W_{\sigma\bar\sigma}\right]^{-1}$, where
we have used the normalization condition $p_\uparrow + p_\downarrow=1$.
Note that since elastic cotunneling leaves the dot state invariant,
besides the spin-flip rate $W^\mathrm{flip}_{\bar\sigma\sigma}$ only
the rates for the inelastic cotunneling processes appear in the master
equation. 

\begin{figure}[ht]
  \centerline{\includegraphics[width=0.48\textwidth]{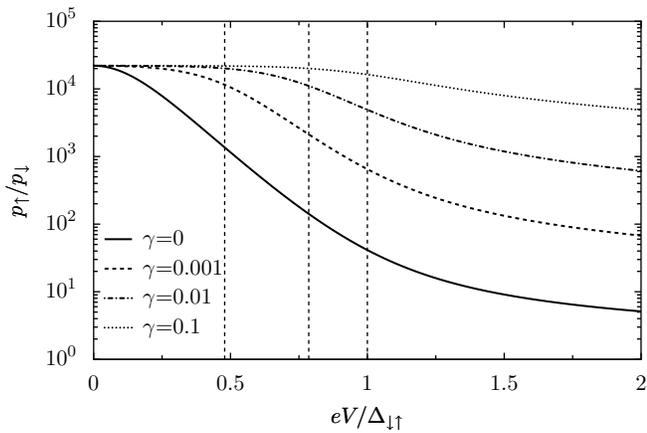}}
  \caption{Ratio $p_\uparrow/p_\downarrow$ of the populations as a
    function of the bias voltage~$V$ in the presence of an Ohmic bath
    with different coupling strengths~$\gamma$. The parameters are
    $\Delta_{\downarrow\uparrow}=0.1 \Escale$, $T=0.01 \Escale$ and
    $\hbar\Gamma_\mathrm{L}=\hbar\Gamma_\mathrm{R}=0.02\Escale$. Here,
    $\Escale=|E_1-\mu|$ is the energy difference between the energy
    $E_1=E_\uparrow=E_\downarrow$ of the degenerate dot level and the
    Fermi energy~$\mu=\mu_\mathrm{L}=\mu_\mathrm{R}$ in the leads in
    the absence of both the magnetic field and the bias voltage. The
    dashed vertical lines indicate the onset of the heating regime
    defined by $W^\mathrm{flip}_{\uparrow\downarrow} =
    W^\mathrm{cot}_{\downarrow\uparrow}$.}
  \label{fig:populations}
\end{figure}

Figure~\ref{fig:populations} shows the ratio $p_\uparrow/p_\downarrow$ of the
populations as a function of the bias voltage~$V$ in the presence of
an Ohmic environment for different coupling strengths~$\gamma$.  Near
thermal equilibrium, i.e., for voltages far below the onset of
inelastic cotunneling, i.e., $eV \ll |\Delta_{\downarrow\uparrow}|, k_\mathrm{B}T$,
the total rates and thus the populations fulfil the detailed balance
relation $p_{\bar\sigma}/p_\sigma =
W_{\bar\sigma\sigma}/W_{\sigma\bar\sigma}=
\exp(-\Delta_{\bar\sigma\sigma}/k_\mathrm{B}T)$. For higher bias
voltages, inelastic cotunneling leads to a population of the excited
dot state, i.e., a heating of the dot.  This heating effect becomes
less pronounced in the presence of the spin-phonon coupling, where the
relaxation due to the rate~\eqref{eq:flip} equilibrates the system.
For a quantitative estimate of the heating effect, we consider the
deviation~$h$ of the population ratio from its equilibrium value:
\begin{equation}
  \label{eq:8}
  h = \left|\frac{p_\uparrow}{p_\downarrow} -
  \exp(\Delta_{\downarrow\uparrow}/k_\mathrm{B}T)\right|
  \,.
\end{equation}
Using the detailed balance relation for the rates
$W^\mathrm{flip}_{\bar\sigma\sigma}$, we then find
\begin{equation}
  \label{eq:9}
  h  {}= \frac{|\exp(\Delta_{\downarrow\uparrow}/k_\mathrm{B}T) -
    W^\mathrm{cot}_{\uparrow\downarrow}/W^\mathrm{cot}_{\downarrow\uparrow} |}
  {1+
    W^\mathrm{flip}_{\uparrow\downarrow}/W^\mathrm{cot}_{\downarrow\uparrow}}
  \,.
\end{equation}
Thus, even for a non-zero difference in the numerator, the heating
effect becomes suppressed for $W^\mathrm{flip}_{\uparrow\downarrow}\gg
W^\mathrm{cot}_{\downarrow\uparrow}$. In Fig.~\ref{fig:populations},
we indicate the onset voltages of the heating regime, defined by the
relation $W^\mathrm{flip}_{\uparrow\downarrow} =
W^\mathrm{cot}_{\downarrow\uparrow}$ by vertical dashed lines.

Finally, we remark that heating effects are suppressed when the
dot-lead coupling is asymmetric,~\cite{Golovach2004b} say
$\Gamma_\mathrm{L}\gg\Gamma_\mathrm{R}$. In this case the excitation
of the dot, which is dominated by processes involving the tunneling
from one lead to the other and which, hence, is proportional to the
product $\Gamma_\mathrm{L} \Gamma_\mathrm{R}$, is negligible compared
to the relaxing effects resulting from the coupling to the left lead,
which are proportional to $\Gamma_\mathrm{L}^2$. Such a situation may
occur, e.g., when measuring the current through an atom or molecule on
a surface which is contacted on one side by an STM
tip~\cite{Heinrich2004a} (cf.\ also discussion at the end of
Sec.~\ref{sec:kogan}).

\section{Differential conductance}

\label{sec:conductance}

Having all the information for the evaluation of the
current~\eqref{eq:current2} at hand, we can now turn to the
experimentally relevant quantity, the differential conductance~$g = \d
I/\d V$. As a function of the bias voltage, this quantity shows a
characteristic step at the onset of inelastic cotunneling around
$|\Delta_{\downarrow\uparrow}|/e$, when another transport channel
becomes accessible. It is known that the width of this step is
proportional to the temperature $T$, while the broadening of the dot
states due to their coupling to the leads does not play a role.
Compared to a measurement in the sequential tunneling regime, an
experiment can thus achieve a much higher precision for the
determination of the splitting~$\Delta_{\downarrow\uparrow}$ (see
Refs.~\onlinecite{Heinrich2004a} and \onlinecite{Kogan2004a}).  An open question,
however, is how spin-flip processes taking place during the
cotunneling process affect such a measurement.

\begin{figure}[ht]
  \centerline{\includegraphics[width=0.95\linewidth]{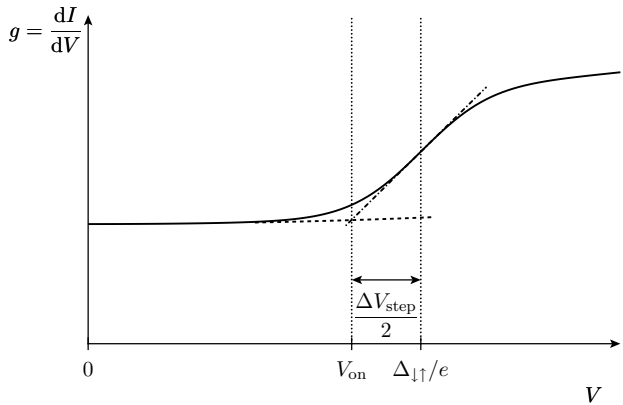}}
  \caption{Sketch of the procedure employed for the determination of
    the onset $V_\mathrm{on}$ and of the width~$\Delta
    V_\mathrm{step}$ of the inelastic-cotunneling step. The dashed
    line shows the extrapolation of the differential conductance curve below the
    step. The dash-dotted line corresponds to the linear approximation
    of the step shape.}
  \label{fig:width}
\end{figure}

Let us first consider the zero-temperature case without heating where
the dot is always in its ground state which we assume to be the
spin-up state.  Evaluating the current using the approximate
cotunneling rates~\eqref{eq:Icot2} and \eqref{eq:Wcotphinel3}, we then
find for the contribution due to elastic cotunneling
\begin{multline}
  \label{eq:condcotel}
  \!\!\!
  g_\mathrm{el} =
    \frac{e^2\hbar
      \Gamma_\mathrm{L}\Gamma_\mathrm{R}
      }{2\pi\Delta_{\mathrm{L}\uparrow\mathrm{R}\uparrow}^2}
    \bigg\{
    \!
    1
    +
    \frac{1}{4\Delta_{\mathrm{L}\uparrow\mathrm{R}\uparrow}^2} 
    \!
    \left[ 4\pi^2
      (k_\mathrm{B} T)^2 + 3(eV)^2\right]
    \\
    +
    \frac{\gamma(\hbar\omega_\mathrm{c})^{1-s}}{4\pi}
    \left(
    \frac{\Delta_{\mathrm{L}\uparrow\mathrm{R}\uparrow}^2}{\Delta_{\downarrow\mathrm{L}}^2\, \Delta_{\uparrow\mathrm{R}}^2}
    +
    \frac{\Delta_{\mathrm{L}\uparrow\mathrm{R}\uparrow}^2}{\Delta_{\uparrow\mathrm{L}}^2\,      \Delta_{\downarrow\mathrm{R}}^2}
  \right)
  \\\times
  (k_\mathrm{B}T)^{s}
  \left[
  k_\mathrm{B} T
  J_s\left(eV/k_\mathrm{B}T\right)
  +
  eV J'_s\left(eV/k_\mathrm{B}T\right)
  \right]\!\!
  \bigg\}\,,
\end{multline}
where $J'_s(a)= \d J_s(a)/\d a$. In order to obtain a compact
expression for the inelastic cotunneling current, we furthermore
assume positive voltages $V>0$ and not too high temperatures
$k_\mathrm{B}T\ll\Delta_{\downarrow\uparrow}$. Then only the tunneling
from the left to the right contact is relevant, i.e., the total
current $I$ is to a very good approximation given by
$I_{\mathrm{R}\mathrm{L}}$.  Furthermore, away from the step around
$V\approx\Delta_{\downarrow\uparrow}/e$, we can use that up to
exponentially small corrections $\Theta(\epsilon)\approx0$ for
$\epsilon\ll k_\mathrm{B}T$ and $\Theta(\epsilon)\approx\epsilon$ for
$\epsilon\gg k_\mathrm{B}T$.
Thus, for $eV \ll \Delta_{\downarrow\uparrow}$, the inelastic
cotunneling current vanishes and for voltages
$eV\gg\Delta_{\downarrow\uparrow}$, it contributes with
\begin{widetext}
\begin{multline}
  \label{eq:condcotinel}
  g_\mathrm{inel} = 
  \frac{e^2\hbar
    \Gamma_\mathrm{L}\Gamma_\mathrm{R}
      }{2\pi\Delta_{\mathrm{L}\downarrow\mathrm{R}\uparrow}^2}
    \bigg\{
    1
    +         
    \frac{1}{4\Delta_{\mathrm{L}\downarrow\mathrm{R}\uparrow}^2} 
    \left[ 4\pi^2
      (k_\mathrm{B} T)^2 + 3 \left(eV-\Delta_{\downarrow\uparrow}\right)^2\right]
    \\
    +
    \frac{\gamma(\hbar\omega_\mathrm{c})^{1-s}}{4\pi}
    \left(
    \frac{\Delta_{\mathrm{L}\downarrow\mathrm{R}\uparrow}^2}{\Delta_{\uparrow\mathrm{L}}^2\, \Delta_{\uparrow\mathrm{R}}^2}
    +
    \frac{\Delta_{\mathrm{L}\downarrow\mathrm{R}\uparrow}^2}{\Delta_{\downarrow\mathrm{L}}^2\,
      \Delta_{\downarrow\mathrm{R}}^2}
  \right)
    (k_\mathrm{B}T)^{s}
    \big[
      k_\mathrm{B}T
      J_s\left(\left(eV-\Delta_{\downarrow\uparrow}\right)/k_\mathrm{B}T\right)
      +
      \left(eV-\Delta_{\downarrow\uparrow}\right)
      J'_s\left(\left(eV-\Delta_{\downarrow\uparrow}\right)/k_\mathrm{B}T\right)
  \big]
  \bigg\}
\end{multline}
\end{widetext}
to the differential conductance. Particularly interesting is the step in between,
which is centered around $V =\Delta_{\downarrow\uparrow}/e$. In order
to obtain an estimate for its width $\Delta V_\mathrm{step}$, we
determine the onset $V_\mathrm{on}$ of inelastic cotunneling by using
a linear approximation of the differential conductance around the center of the
step, $g(V)\approx g(\Delta_{\downarrow\uparrow}/e) +
g'(\Delta_{\downarrow\uparrow}/e)\, (V-\Delta_{\downarrow\uparrow}/e)$
and intersecting with the purely elastic cotunneling
curve~\eqref{eq:condcotel} (see
Fig.~\ref{fig:width}): $g_\mathrm{el}(V_\mathrm{on}) =
g(\Delta_{\downarrow\uparrow}/e) + g'(\Delta_{\downarrow\uparrow}/e)\,
(V_\mathrm{on}-\Delta_{\downarrow\uparrow}/e)$. Approximately, we can determine the solution of this non-linear
equation by employing a linear approximation for the elastic cotunneling
curve around the center of the step. Doing so, we obtain $V_\mathrm{on}
= \Delta_{\downarrow\uparrow}/e -
g_\mathrm{inel}(\Delta_{\downarrow\uparrow}/e)/g'_\mathrm{inel}(\Delta_{\downarrow\uparrow}/e)$.
Defining the width as twice the difference $V_\mathrm{on}
- \Delta_{\downarrow\uparrow}/e$ gives
\begin{multline}
  \label{eq:deltaVstep}
  \Delta V_\mathrm{step} =
  \frac{6 \, k_\mathrm{B}T}{e}
  \bigg\{
    1 -
    3
    \left(
      \frac{k_\mathrm{B}T}{\Delta_{\mathrm{L}\downarrow\mathrm{R}\uparrow}}
    \right)^2
    \bigg[
    1 
    +
    \gamma\,
    \frac{1}{6\pi}
    \\
    \times
    \left(
    \frac{\Delta_{\mathrm{L}\downarrow\mathrm{R}\uparrow}^4}{\Delta_{\uparrow\mathrm{L}}^2\, \Delta_{\uparrow\mathrm{R}}^2}
    +
    \frac{\Delta_{\mathrm{L}\downarrow\mathrm{R}\uparrow}^4}{\Delta_{\downarrow\mathrm{L}}^2\,
      \Delta_{\downarrow\mathrm{R}}^2}
  \right)
  \\
  \times
  \big(
  3 J_s''(0) + 2 J_s'(0)
  \big)
  \left(\frac{k_\mathrm{B}T}{\hbar\omega_\mathrm{c}}\right)^{s-1}
  \bigg]
  \bigg\}\,.
\end{multline}
As mentioned above, the width is essentially proportional to the
temperature. We also find that the spin-flip processes only modify the
higher-order corrections in
$k_\mathrm{B}T/\Delta_{\mathrm{L}\downarrow\mathrm{R}\uparrow}$.
Here, the spin-flip coupling contribution is proportional to $\gamma$
and to a positive prefactor of the order of one and has a temperature
dependence which depends on the ratio of thermal and cutoff energy.
In fact, we find that with increasing coupling constant $\gamma$ the width of
the inelastic cotunneling step  becomes reduced.

\begin{figure}[ht]
  \centerline{\hbox to 0pt{(a)}\includegraphics[width=0.48\textwidth]{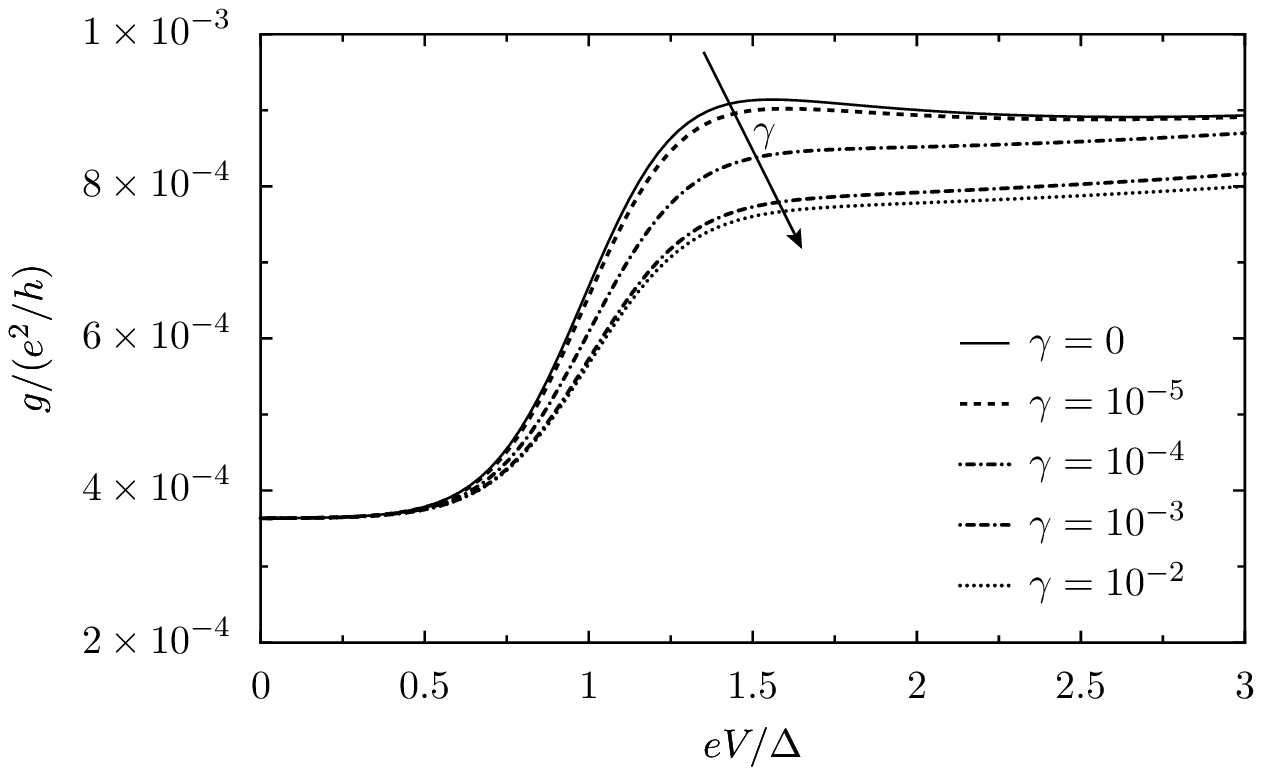}}

  \vspace{0.5cm}
  \centerline{\hbox to 0pt{(b)}\includegraphics[width=0.48\textwidth]{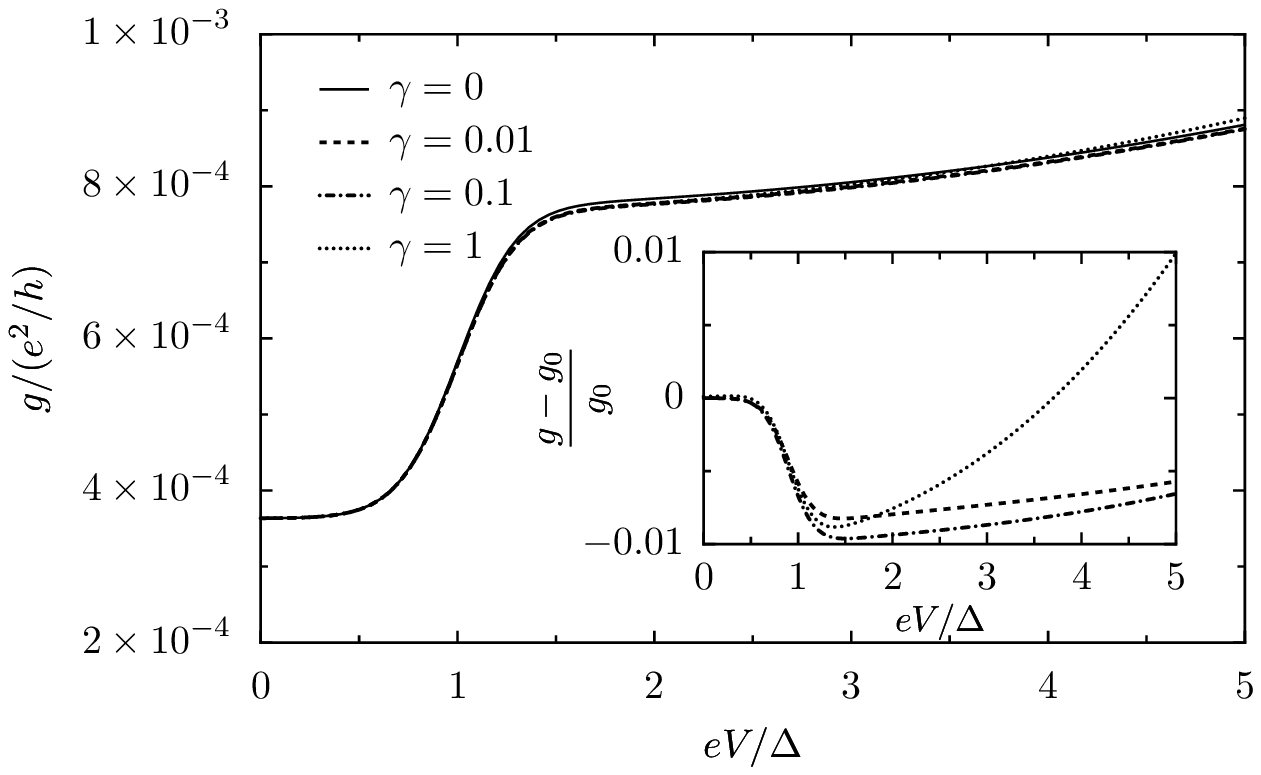}}
  \caption{Cotunneling conductance~$g$ as a function of the bias
    voltage $V$ in the presence of an Ohmic bath with different
    coupling strengths~$\gamma$.  The parameters are
    $\Delta_{\downarrow\uparrow}=0.1 \Escale$, $T=0.01 \Escale$ and
    (a) $\hbar\Gamma_\mathrm{L}=\hbar\Gamma_\mathrm{R}=0.02\Escale$,
    (b) $\hbar\Gamma_\mathrm{L} = 0.2\Escale$ and
    $\hbar\Gamma_\mathrm{L} = 0.002 \Escale$ with $E_0$ as defined in
    the caption of Fig.~\ref{fig:populations}.  In the inset of
    panel~(b), the relative difference between the differential conductance~$g$ in
    the presence of the spin-phonon coupling and the differential conductance~$g_0$
    without spin-phonon coupling is depicted. }
  \label{fig:conductance_V}
\end{figure}

We now turn to the discussion of the heating effects by taking into
account the stationary probability distribution according to the
master equation~\eqref{eq:mastereq}. Figure~\ref{fig:conductance_V}
shows the differential conductance as a function of the voltage in the presence of
an Ohmic spin-flip coupling of strength~$\gamma$. In the upper panel,
the case of a symmetric dot-lead coupling is depicted. As discussed
above, this is the situation where heating effects are most relevant,
and for $\gamma=0$ we indeed observe a noticeable deviation from the
step-like behavior for voltages $V\gtrsim
\Delta_{\downarrow\uparrow}/e$. With increasing spin-flip coupling
strength~$\gamma$, however, the dot is driven to equilibrium and the
overshooting of the differential conductance curve disappears. As mentioned at the end of
Sec.~\ref{sec:master-equation}, heating effects are suppressed when
the dot-lead coupling is strongly asymmetric. Such a situation is
shown in panel~(b) of Fig.~\ref{fig:conductance_V}. In this case, the
influence of the spin-phonon coupling is much weaker. For larger
voltages, the effect of phonon-assisted inelastic cotunneling becomes
visible, as can be inferred from the inset of panel~(b).

\begin{figure}[ht]
  \centerline{\includegraphics[width=0.48\textwidth]{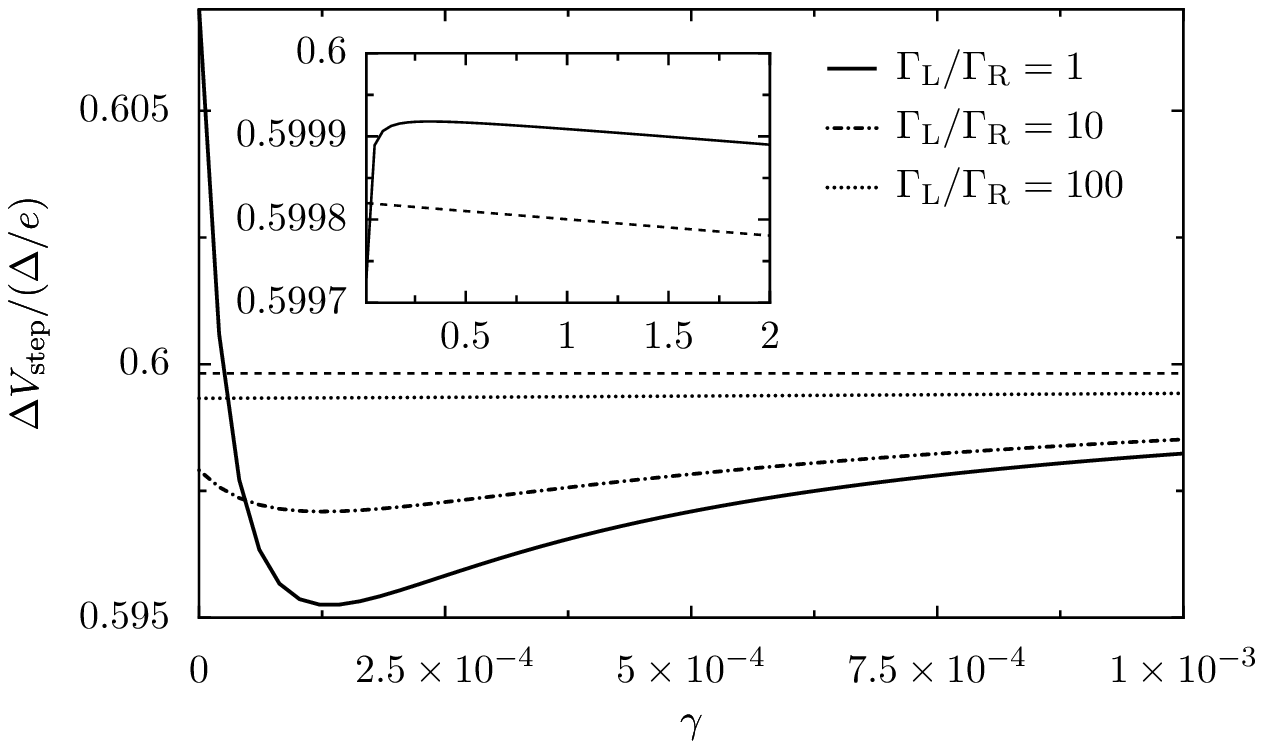}}
  \caption{Step width $\Delta V_\mathrm{step} =
    2g'_\mathrm{inel}(\Delta_{\downarrow\uparrow}/e)/g_\mathrm{inel}(\Delta_{\downarrow\uparrow}/e)$
    as a function of the coupling strength~$\gamma$ to an Ohmic bath
    for different ratios of the tunneling rates $\Gamma_\mathrm{L,R}$,
    where the product $\hbar^2\Gamma_\mathrm{L}\Gamma_\mathrm{R}=(0.02
    \Escale)^2$ is kept constant.  The other parameters are like in
    Fig.~\ref{fig:conductance_V}. The horizontal dashed line shows the
    width according to Eq.~\eqref{eq:deltaVstep} for $\gamma=0$. In
    the inset, the behavior for larger $\gamma$ is shown together with
    the step width without heating according to
    Eq.~\eqref{eq:deltaVstep} (dashed).}
  \label{fig:stepwidth}
\end{figure}

In Fig.~\ref{fig:stepwidth}, the step-width defined by the onset of
inelastic cotunneling is shown as a function of the spin-flip coupling
strength~$\gamma$ for different ratios
$\Gamma_\mathrm{L}/\Gamma_\mathrm{R}$. For equal coupling to both
leads, the main effect again comes from the suppression of the heating
with increasing $\gamma$. The inset of Fig.~\ref{fig:stepwidth}
depicts the $\gamma$-dependence for larger $\gamma$. As described by
the analytical expression~\eqref{eq:deltaVstep} the width drops
linearly. Due to the finite temperature, there is a slight offset
compared to the full solution, however.

\section{Comparison with experiments}

\label{sec:kogan}

We now apply the general results to the specific experimental
situation of Ref.~\onlinecite{Kogan2004a} and analyze the relevance of
spin-phonon coupling effects in this case. In order to do so, we model
the 2D quantum dot by an isotropic parabolic lateral
confinement characterized by the frequency $\omega_0$ in the absence
of a magnetic field. In a magnetic field $B_z$ perpendicular to the
two-dimensional electron gas (2DEG), the relevant frequencies then are $\omega_{1,2} = \Omega
\mp\omega_\mathrm{c}/2$ with the cyclotron frequency
$\omega_\mathrm{c} = |e| B_z / m^\ast c$, where $m^\ast$ is the effective
electron mass and $\Omega = \sqrt{\omega_0^2+\omega_\mathrm{c}^2/4}$
corresponding to the effective lateral confinement length
$l=\sqrt{\hbar m^\ast/\Omega}$ (for more details, we refer the reader
to Ref.~\onlinecite{Bulaev2005a}).
The parameters for the GaAs quantum dot of Ref.~\onlinecite{Kogan2004a} are
$g=-0.16$, $m^\ast=0.067m_\mathrm{e}$, $B_z=11$T and we assume $\hbar\omega_0=1.1$meV.

We consider the influence of both Rashba~\cite{Bychkov1984a} and
Dresselhaus~\cite{Dresselhaus1955a} spin-orbit coupling, which are
characterized by spin-orbit lengths
$\lambda_\mathrm{R}=\lambda_\mathrm{D}=8\mu$m (see Appendix A). For a piezoelectric
coupling to longitudinal and transversal phonon modes in the limit
$\omega\ll c_\alpha \min( \ell, d)$, where $d\approx 5$nm is the
vertical confinement length, the spectral
density~\eqref{eq:spectraldens} of the spin-phonon coupling is cubic, i.e.,
$s=s_\mathrm{PE} = 3$. The prefactors for the longitudinal and
transverse contributions are given by
\begin{align}
  \gamma_\mathrm{PE,L} & =
  \frac{Q_\mathrm{D}^2 + Q_\mathrm{R}^2}{70\pi \hbar
    \,\varrho\, c_\mathrm{L}^3}
  \left(\frac{e h_{14}}{\kappa}\right)^2
  \\
  \gamma_\mathrm{PE,T} & =
  \frac{Q_\mathrm{D}^2 + Q_\mathrm{R}^2}{105\pi \hbar
    \,\varrho \,c_\mathrm{T}^3}
    \left(\frac{e h_{14}}{\kappa}\right)^2
\end{align}
with the speed of sound $c_\mathrm{L}=4.73 \times 10^5\,$cm/s and
$c_\mathrm{T}=3.35\times 10^5\,$cm/s of the longitudinal and transversal
acoustic phonon modes, respectively. The density of GaAs is
$\varrho=5.319\,$g/cm$^3$ and the electron-phonon coupling parameters
are $eh_{14}=1.2\times10^7\,$eV/cm and $\kappa=13.2$.  The spin-orbit
coupling strength enters via the dimensionless coupling
constants~\cite{Golovach2004a, Bulaev2005a}
\begin{equation}
  \label{eq:couplconst}
  Q_\mathrm{R,D} =
  \frac{l}{\lambda_\mathrm{R,D}}
  \left[-\frac{\hbar \omega_1}{\hbar\omega_1\mp\Delta_{\downarrow\uparrow}} +
  \frac{\hbar\omega_2}{\hbar\omega_2\pm\Delta_{\downarrow\uparrow}}\right] 
  \,.
\end{equation}
The cut-off frequencies for the piezo-electric coupling are given by
$\omega_\mathrm{c,PE,L} = c_\mathrm{L}/l$ and $\omega_\mathrm{c,PE,T}
= c_\mathrm{T}/l$.  Note that the leading contribution to
$Q_\mathrm{R,D}$ is linear in $B_z$, and consequently the
rate~\eqref{eq:flip} is proportional to $B_z^5$ for low magnetic
fields~$B_z$.

In the case of a deformation potential coupling to acoustic phonons,
the spectral density is proportional to~$\omega^5$, i.e.,
$s=s_\mathrm{DA}=5$, and the prefactor is given by
\begin{equation}
  \label{eq:3}
  \gamma_\mathrm{DA} =
  \frac{  \Xi_0^2\, \left(Q_\mathrm{D}^2 +
      Q_\mathrm{R}^2\right)}{24\pi \hbar\,\varrho \,l^2\,c_\mathrm{L}^3}\,
\end{equation}
with $\Xi_0=6.7\,$eV. The cut-off is at the frequency
$\omega_\mathrm{c,DA} = c_\mathrm{L}/l$. Taking into account the
magnetic-field-dependence of the prefactor~$\gamma_\mathrm{DA}$, the
leading contribution to the rate~\eqref{eq:flip} is proportional to
$B_z^7$ at low magnetic fields. Thus, the spin-phonon coupling due to
the deformation potential electron-phonon interaction is only relevant
for rather high magnetic fields.  Furthermore, we take a temperature
$T=15\,$mK and dot-lead coupling rates
$\hbar\Gamma_\mathrm{L}=\hbar\Gamma_\mathrm{R}=17.5\,\mu$eV from
Ref.~\onlinecite{Kogan2004a}. Finally, we assume a value
$|E_1-\mu|=0.4$meV, where $E_1$ and $\mu$ are the energy of the
degenerate dot-level and the Fermi energy, respectively, in the
absence of the bias voltage and the magnetic field. An external bias
is assumed to drop symmetrically across both contacts.

\begin{figure}[ht]
  \centerline{\includegraphics[width=0.48\textwidth]{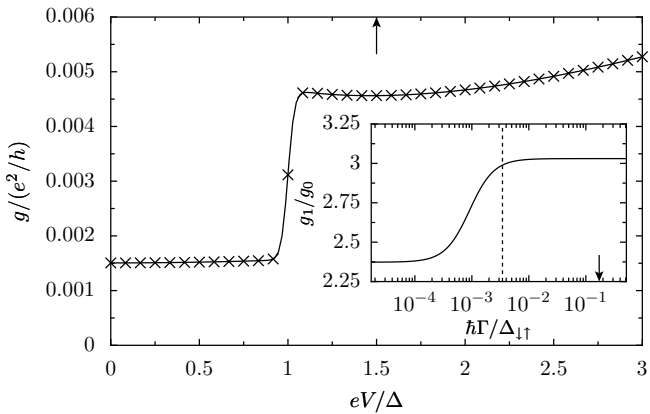}}
  \caption{Differential conductance~$g$ as a function of the bias
    voltage~$V$ for the parameters given in the main text. The crosses
    show the result in the absence of the spin-phonon coupling. In the
    inset, the ratio of the differential conductance $g_1$ at
    $V_1=1.5\Delta_{\downarrow\uparrow}/e$ (see arrow in the main
    panel) and the differential conductance $g_0$ at zero voltage for different
    values of the dot-lead coupling rates
    $\Gamma=\Gamma_\mathrm{L}=\Gamma_\mathrm{R}$. The dashed vertical
    line designates the value given by Eq.~\eqref{eq:Gammatransition}
    and the arrow the dot-lead coupling rate in the
    experiment~\cite{Kogan2004a}.}
  \label{fig:conductance_V_goldhaber}
\end{figure}

Figure~\ref{fig:conductance_V_goldhaber} depicts the differential conductance as a
function of the bias voltage for the given set of parameters.
Comparing the result of the calculation including the spin-orbit
induced spin-phonon coupling (solid line) with the one without this
coupling (crosses), we can conclude that these effects play a
negligible role in the experiment by Kogan et al. However, as shown in
the inset of Fig.~\ref{fig:conductance_V_goldhaber}, by reducing the
dot-lead coupling by a bit more than one order of magnitude, one
reaches a regime where the equilibration of the dot due to the
spin-phonon coupling becomes relevant. As discussed in
Sec.~\ref{sec:master-equation}, the transition to this regime is
determined by the relation $W^\mathrm{flip}_{\uparrow\downarrow} =
W^\mathrm{cot}_{\downarrow\uparrow}$, which for
$eV\gtrsim\Delta_{\downarrow\uparrow}$ yields the tunneling rate
\begin{equation}
  \label{eq:Gammatransition}
  \Gammatransition
  =
  |\Delta_{\mathrm{L}\downarrow\mathrm{R}\uparrow}|
  \sqrt{\frac{2\pi \,W^\mathrm{flip}_{\uparrow\downarrow}}{\hbar\Theta(eV-\Delta_{\downarrow\uparrow})}}\,.
\end{equation}
This relation can be employed for the determination of the spin-flip
rate~$W^\mathrm{flip}_{\uparrow\downarrow}$ by means of a cotunneling
transport measurement. In order to do so, one has to determine the
critical tunneling rate~$\Gammatransition$ marking the transition
between the regime of the dot being in equilibrium and the one where
heating effects are relevant. As an indicator for this transition one
can measure the ratio of the differential conductance above and below the inelastic
cotunneling step. Determining this quantity as a function of the
dot-lead coupling strength yields $\Gammatransition$ (cf. inset of
Fig.~\ref{fig:conductance_V_goldhaber}) and via
Eq.~\eqref{eq:Gammatransition} the spin-flip rate
$W^\mathrm{flip}_{\uparrow\downarrow}$.

Let us finally discuss whether such a measurement scheme would also
apply to transport experiments through single atoms or molecules on a
surface, e.g., the ones reported in Ref.~\onlinecite{Heinrich2004a}.
There, the tunnel-coupling to the two leads is strongly asymmetric,
one contact being via the insulating surface and the other via an STM
tip.  As has been discussed at the end of
Sec.~\ref{sec:master-equation}, heating effects are suppressed in this
case. Indeed, the form of the differential conductance-voltage
characteristics seen in Ref.~\onlinecite{Heinrich2004a} can---above
the Kondo temperature---be very well described by an equilibrium
probability distribution of the spin state of the atom.  However, for
a not too small tunnel-coupling across the insulating layer, it
should be possible to enter the heating regime by adjusting the
atom-tip coupling strength appropriately. If this coupling strength is
not too far away from the critical value given by
Eq.~\eqref{eq:Gammatransition}, it should be possible to measure the
intrinsic spin-relaxation rate due to phonons by using a similar
scheme as the one proposed above, although only the atom-STM coupling
strength can be controlled easily. On the other hand, it is in
principle feasible to control the molecule-surface coupling strength
by modifying the chemical binding of the molecule to the surface, as
was shown in Ref.~\onlinecite{Zhao2005a}, so the proposed scheme could
be applied to such a situation, as well.

\section{Scaling behavior}

\label{sec:pms}

So far, we have restricted the discussion to the case of temperatures
well above the so-called Kondo-temperature $T_\mathrm{K}$. For lower
temperatures, correlations between between the dot spin and a
collective spin degree of freedom of the electrons in the leads become
important and a perturbative treatment of the problem breaks down.
This leads to the so-called Kondo effect, which has been analyzed in
great detail in the literature for the Anderson
model~\cite{Hewson1993a}. We now briefly discuss the effect of
additional spin-flip processes due to the term $H_\B$ using the poor man's
scaling approach put forward by Anderson~\cite{Anderson1970a}.  In
order to keep the discussion as simple as possible, we consider a
simplified version of our effective Hamiltonian \eqref{eq:SWT}, which
ignores the lead-, momentum- and spin-dependence of the coupling
constants as well as the direct interaction between the lead states:
\begin{equation}
  \label{eq:33}
  \begin{split}
    \bar H = {} 
    & 
    H_0 -
    J \mathbf{S}_\mathrm{d} \cdot \mathbf{s}
    + J_\B \left(X S^x_\mathrm{d} + Y S^y_\mathrm{d}\right)
    \\& 
  - J_1 \left(X S^x_\mathrm{d} + Y S^y_\mathrm{d} \right) n(0)
  + J_2 \left(X  s^x + Y s^y \right)
  \end{split}
\end{equation}
Here, we have introduced the spin operators 
\begin{align}
  \mathbf{s} & {}= \frac{1}{2} 
  \sum_{\ell \ell'  \mathbf{k} \mathbf{k}'\sigma \sigma'}
  c^\dagger_{\ell\mathbf{k}\sigma}
  \, \boldsymbol{\tau}_{\sigma\sigma'}\,
  c_{\ell'\mathbf{k}'\sigma'} 
  \text{ and}
  \\
  \mathbf{S}_\mathrm{d} & {}= \frac{1}{2} \sum_{\sigma \sigma'}
  d^\dagger_{\sigma}\, \boldsymbol{\tau}_{\sigma\sigma'}\,
  d_{\sigma'}\,,
\end{align}
where $\boldsymbol{\tau}$ is the vector consisting of the three Pauli
matrices.  The density of lead electrons at the position of the dot is
given by
\begin{equation}
  n(0) = \sum_{\ell\ell'\mathbf{k} \mathbf{k}'\sigma}
  c^\dagger_{\ell\mathbf{k}\sigma} c_{\ell'\mathbf{k}'\sigma}\,.
\end{equation}
The phonon operators are defined as $X=\sum_\mathbf{q}
M_{\mathbf{q},x}(a_\mathbf{q}+a^\dagger_{-\mathbf{q}})$ and
$Y=\sum_\mathbf{q}
M_{\mathbf{q},y}(a_\mathbf{q}+a^\dagger_{-\mathbf{q}})$.

During the scaling procedure, contributions due to excitations of
electrons and holes far away from the Fermi energy $\mu=0$, at a
cutoff energy $\pm D$, are successively integrated out. If the
resulting effective interaction is again of the form~\eqref{eq:33}, we
can absorb it in the coupling constants.  Note that in general, the
coupling constants thereby acquire a momentum dependence.  However, as
long as the cutoff energy is still large enough, this dependence can
be ignored for the relevant states near the Fermi energy. Doing so, we
obtain the usual scaling equations for the exchange constant~$J$:
\begin{equation}
  \label{eq:35}
  \frac{\d J}{\d D} = -2 \nu \frac{J^2}{D}\,,
\end{equation}
which, in particular, is not influenced by the other terms in the
Hamiltonian~\eqref{eq:33}.  The renormalization of the
electron-phonon coupling constant is given by
\begin{equation}
  \label{eq:36}
  \frac{\d J_\B}{\d D} = 
  \nu {J J_2}
  \sum_{\ell\mathbf{k}}^{|\epsilon_{\ell\mathbf{k}}|<D}\frac{1}{D+|\epsilon_{\ell\mathbf{k}}|}
  = 
  (4 \ln 2)\,\,
  \nu^2 {J J_2} 
\end{equation}
while the coupling constants $J_1$ and $J_2$ remain invariant. Here,
$\nu$ is the density of states in the leads. In the anti-ferromagnetic
case, where $J>0$ scales to $\infty$ as $D\to0$, the same holds true
for the spin-phonon coupling constant~$J_\B$. We conclude that far
away from the active region around the Fermi energy, neither the pure
spin-flip term proportional to $J_\B$ nor the phonon-assisted
tunneling terms, i.e., the last two terms in the
Hamiltonian~\eqref{eq:33}, modify the scaling behavior of the
Kondo-type coupling of the dot spin to the leads.

\section{Conclusions}
\label{sec:conclusions}

We have theoretically studied the influence of a spin-phonon coupling
on the cotunneling through a nanoscale system. By means of a
Schrieffer-Wolff transformation, we have derived an explicit
Hamiltonian in which the relevant cotunneling processes appear to
lowest order. This allowed us to evaluate the rates for elastic and
inelastic cotunneling in the presence of spin-flip processes. We found
that the width of the inelastic cotunneling step is only weakly
influenced by the spin-flip processes.  More important are the
relaxation effects of the spin-phonon coupling which counteract the
heating of the quantum dot due to inelastic cotunneling. Considering a
realistic model for a recent experiment~\cite{Kogan2004a}, we propose
a new way of determining the spin relaxation rate due to spin-phonon
coupling in a cotunneling experiment.

\begin{acknowledgments}
  The authors thank D.~Bulaev, W.A.~Coish, V.N.~Golovach, F.~von~Oppen,
  D.~Saraga, and
  P.~Simon for helpful discussion. Financial support by the EU RTN QuEMolNa, the NCCR
  Nanoscience, the Swiss NSF, DARPA, ARO, and ONR is acknowledged.
\end{acknowledgments}

\appendix

\section{Derivation of spin-phonon coupling Hamiltonian}
\label{sec:deriv-spin-phon}

In this appendix, we derive the spin-phonon coupling~\eqref{eq:HB}
starting from a Hamiltonian describing an electron in a
two-dimensional electron gas with an external magnetic field $B_z$ in
z-direction
\begin{equation}
  \label{eq:a1}
  H_0 =
  \frac{\mathbf{P}^2}{2m^\ast} + U(\mathbf{r}) +
  \frac{1}{2} g\mu_\mathrm{B} B_z \sigma_z
\end{equation}
where $\mathbf{P}=\mathbf{p}+(|e|/c)\mathbf{A}(\mathbf{r})$ is the
kinetic momentum, $U(\mathbf{r})$ is the confinement potential, and
$\mathbf{A}(\mathbf{r})= (B_z/2)(-y,x,0)$ is the vector potential. For
$x,y,z$ pointing along the main crystallographic axes of the GaAs
crystal, and the electron gas lying in the $[001]$ plane, the spin-orbit
interaction assumes the form 
\begin{equation}
  \label{eq:a2}
  H_\mathrm{SO} = H_\mathrm{D} + H_\mathrm{R}
\end{equation}
with the Dresselhaus~\cite{Dresselhaus1955a} and Rashba~\cite{Bychkov1984a} contributions
\begin{equation}
  \label{eq:a3}
  H_\mathrm{D} = \beta_\mathrm{D}(-\sigma_x P_x + \sigma_y P_y)\,,\
  H_\mathrm{R} = \alpha_\mathrm{R}(\sigma_x P_y - \sigma_y P_x)\,.
\end{equation}
The strength of the spin-orbit coupling is conveniently measured in
terms of the spin-orbit lengths $\lambda_\mathrm{D}=\hbar/m^\ast\beta_\mathrm{D}$
and $\lambda_\mathrm{R}=\hbar/m^\ast\alpha_\mathrm{R}$.  Finally, the orbital
degrees of freedom of the electron are coupled to the phonon system
via the electron-phonon interaction $H_\mathrm{e-ph}= \sum_\mathbf{q}
M_\mathbf{q}(\mathbf{r}) (a_\mathbf{q} + a^\dagger_{-\mathbf{q}})$,
where a branch index has been suppressed.

Applying to the Hamiltonian $H = H_0 + H_\mathrm{SO} +
H_\mathrm{e-ph}$ a Schrieffer-Wolff transformation which eliminates
$H_\mathrm{SO}$ to lowest order and projecting on the orbital ground
state, we obtain to lowest order the \textit{spin-phonon} coupling
\begin{equation}
  H_\B = \langle 0|[S, H_\mathrm{e-ph}]|0\rangle
\end{equation}
where the generator $S$ of the canonical transformation has to fulfil
$[S,H_0]=-H_\mathrm{SO}$. From Eqs.~\eqref{eq:a1}--\eqref{eq:a3}, we
can infer that the generator can be written in the form $S = i(A_x
\sigma_x + A_y \sigma_y)$ with $A_x$ and
$A_y$ being Hermitian operators which act on the orbital degrees of freedom
only. Hence, we obtain
\begin{equation}
  H_\B = i \langle 0|[A_x, H_\mathrm{e-ph}]|0\rangle \, \sigma_x +i 
  \langle 0|[A_y, H_\mathrm{e-ph}]|0\rangle \, \sigma_y\,.
\end{equation}
\\
Letting $M_{\mathbf{q},i} = \langle 0|[\Omega_i,
M_\mathbf{q}(\mathbf{r})]|0\rangle$ for $i=x,y$, we thus arrive at the
coupling Hamiltonian~\eqref{eq:HB}.

For a quantum dot with parabolic lateral confinement (see
Sec.~\ref{sec:kogan}), it is useful to choose coordinates rotated by
$45^\circ$~degrees around the $z$-axis\cite{Golovach2004a}, i.e., we
let $x\to(x+y)/\sqrt{2}$ and $y\to(y-x)/\sqrt{2}$. In this new
coordinate frame, the generator S of the canonical transformation is
defined in terms of
\begin{align}
  \label{eq:Omegax}
  A_x = {} & 
  \phantom{{}+{}}
  \frac{\alpha_\mathrm{R}}{2\Omega} \bigg[
  \omega_1\,
  \frac{m^\ast\Omega y + p_x}{\hbar\omega_1 - \Delta_{\downarrow\uparrow}}
  +
  \omega_2\,
  \frac{m^\ast\Omega y - p_x}{\hbar\omega_2 + \Delta_{\downarrow\uparrow}}
  \bigg]
  \nonumber
  \\ &
  +
  \frac{\beta_\mathrm{D}}{2\Omega} \bigg[
  \omega_1\,
  \frac{m^\ast\Omega y + p_x}{\hbar\omega_1 + \Delta_{\downarrow\uparrow}}
  +
  \omega_2\,
  \frac{m^\ast\Omega y - p_x}{\hbar\omega_2 - \Delta_{\downarrow\uparrow}}
  \bigg]\,,
  \\
  A_y = {} & 
  \phantom{{}+{}}
  \frac{\alpha_\mathrm{R}}{2\Omega} \bigg[
  \omega_1\,
  \frac{-m^\ast \Omega x + p_y}{\hbar\omega_1 - \Delta_{\downarrow\uparrow}}
  +
  \omega_2\,
  \frac{-m^\ast \Omega x - p_y}{\hbar\omega_2 + \Delta_{\downarrow\uparrow}}
  \bigg]
  \nonumber
  \\ &
  +
  \frac{\beta_\mathrm{D}}{2\Omega} \bigg[
  \omega_1\,
  \frac{m^\ast \Omega x - p_y}{\hbar\omega_1 + \Delta_{\downarrow\uparrow}}
  +
  \omega_2\,
  \frac{m^\ast \Omega x + p_y}{\hbar\omega_2 - \Delta_{\downarrow\uparrow}}
  \bigg]\,.
\end{align}
Note that this transformation diverges for magnetic field strengths
with $\hbar\omega_{1,2} \pm \Delta_{\downarrow\uparrow}=0$. A detailed
analysis of the relaxation rates in this case can be found in
Ref.~\onlinecite{Bulaev2005a}. On the other hand, for
$|\Delta_{\downarrow\uparrow}| \ll \hbar\omega_{1,2} $, the leading contribution to the relaxation rates
comes from the terms linear in $\Delta_{\downarrow\uparrow}\propto
B_z$ (see Ref.~\onlinecite{Golovach2004a}).

\bibliographystyle{prsty}

\end{document}